\documentclass[prb,twocolumn,amsmath,amssymb,superscriptaddress]{revtex4}

\usepackage{graphicx}
\usepackage{dcolumn}
\usepackage{bm}

\begin{document}

\title{Electron Spin Resonance Spectroscopy via Relaxation of Solid-State Spin Probes at the Nanoscale}

\author{L. T. Hall}
 \email{lthall@physics.unimelb.edu.au}
 \affiliation{School of Physics, University of Melbourne, Victoria 3010, Australia}%
\author{P. Kehayias}
\affiliation{Department of Physics, University of California, Berkeley, California 94720, USA}%
\author{D. A. Simpson}
 \affiliation{School of Physics, University of Melbourne, Victoria 3010, Australia}%
\author{A. Jarmola}
\affiliation{Department of Physics, University of California, Berkeley, California 94720, USA}%
\author{A. Stacey}
\affiliation{Centre for Quantum Computation and Communication Technology, School of Physics, University of Melbourne, Victoria 3010, Australia}%
\author{D. Budker}
\affiliation{Department of Physics, University of California, Berkeley, California 94720, USA}%
\author{L. C. L. Hollenberg}
 \affiliation{School of Physics, University of Melbourne, Victoria 3010, Australia}%
\affiliation{Centre for Quantum Computation and Communication Technology, School of Physics, University of Melbourne, Victoria 3010, Australia}%


\begin{abstract}
Electron Spin Resonance (ESR) describes a suite of techniques for characterising electronic systems, with applications in physics, materials science, chemistry, and biology. However, the requirement for large electron spin ensembles in conventional ESR techniques limits their spatial resolution. Here we present a method for measuring the ESR spectrum of nanoscale electronic environments by measuring the relaxation time ($T_1$) of an optically addressed single-spin probe as it is systematically tuned into resonance with the target electronic system. As a proof of concept we extract the spectral distribution for the P1 electronic spin bath in diamond using an ensemble of nitrogen-vacancy centres, and demonstrate excellent agreement with theoretical expectations. As the response of each NV spin in this experiment is dominated by a single P1 spin at a mean distance of 2.7\,nm, the extension of this all-optical technique to the single NV case will enable nanoscale ESR spectroscopy of atomic and molecular spin systems.
\end{abstract}


\maketitle


\section{Introduction}

\begin{figure}
  \includegraphics[width=\columnwidth]{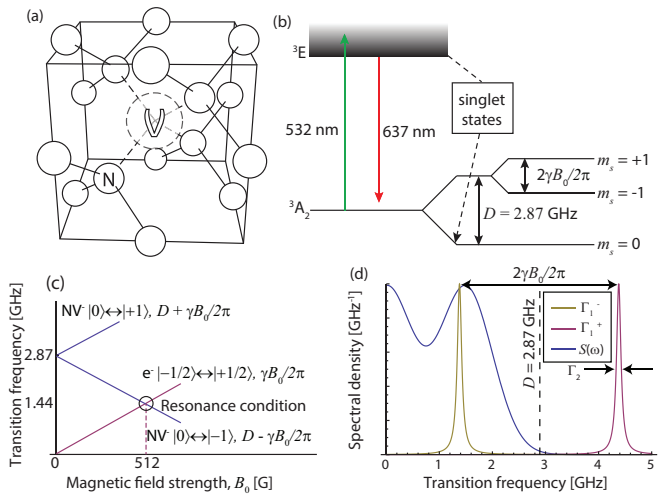}\\
  \caption
  {(a) The nitrogen-vacancy (NV) centre point defect in a diamond lattice,
comprised of a substitutional atomic nitrogen impurity (N) and an adjacent
crystallographic vacancy (V). (b) The NV ground state spin sublevels are separated by $D=2.87$\, GHz. Upon optical excitation at 532 nm, the
population of the $\bigl|0\bigr\rangle$ state may be readout by monitoring the intensity of the emitted red light. (c) Transition frequencies of the NV centre, and that of a single electron as a function of external magnetic field strength, $B_0$. At $B_0=512\,$G the ${|0\rangle\leftrightarrow|-1\rangle}$ transition of the NV is resonant with the electron transition, allowing them to exchange energy. (d) Illustration of the way in which the NV relaxation filter functions,
  $\Gamma_1^+$ and $\Gamma_1^-$ (having width $\Gamma_2 = 1/T_2$), probes specific regions of the environmental
  spectral density. One of the NV transitions is significantly detuned
  from the frequencies exhibited by the environment, and hence cannot be excited. The NV spin thus behaves
  similar to a spin-$\frac{1}{2}$ qubit and the steady state of the NV spin will be an equal mixture
  of only the two states associated with the excited transition, with the third state remaining unpopulated.}\label{RelaxationFilters}
\end{figure}

Techniques to detect electron spin resonance (ESR) have long been used to study materials and systems containing unpaired electron spins, such as metal complexes and organic radicals. From an operational viewpoint, the low spin density and high decay rates of such radicals in-situ, together with the limited sensitivity of ESR detection systems, often makes the task of obtaining an ESR signal above the detection limit difficult. This is further complicated by the fact that a major component of any biological tissue is water, which has a strong electric-dipole mediated absorbtion band in the same region of the electromagnetic spectrum (microwave) as the signals emitted in ESR experiments\cite{Ber96}. As such, there is a great need for a highly sensitive, highly localised technique that may be used to obtain ESR spectra from unpaired electron systems, without the need for microwave control of the sample.

The nitrogen-vacancy (NV) centre in diamond is a promising candidate for such an ideal nanoscale ESR probe. Much attention has been focused on using measurements of NV spin energy sublevels\cite{Deg08,Tay08,Bal08,Maz08} or dephasing rates ($1/T_2$)\cite{Col09,Hal09,Hal10b,McG13} to characterise dynamic processes occurring in external magnetic environments. These protocols have been shown to have remarkable sensitivity to frequencies in the kHz to MHz range, and are thus well suited to
characterising nuclear spin environments\cite{Liu11,Liu12,Sta13,Mam13,Shi14,Dev15}. However, to achieve the
desired sensitivity to the more rapidly fluctuating fields associated with electron spin environments, and more importantly, the
ability to be frequency-selective, complex pulse sequences
are necessary\cite{Cyw08,Hal10a,Bar12}.

Here we focus on on a new, and potentially simpler, technique for extracting the ESR spectrum based on the longitudinal decoherence (or relaxation) time, $T_1$, of the NV centre. Given the GHz transition frequencies of the NV spin in its orbital ground state, owing to its zero-field spitting of $D=2.87\,\mathrm{GHz}$ (Fig.\,1), paramagnetic spins can enhance the longitudinal decay rate of the NV spin, enabling one to detect their presence by monitoring changes in $T_1$\cite{Ste13,Kau13,Tet13,Sus14,Pel14,Kol15}. Magnetic sensing using relaxation times can be much more sensitive for detection because $T_1$ times of an NV centre can be up to three orders of magnitude longer than spin-echo based $T_2$ times\cite{Ste13}.

In the presence of a controlled external magnetic field, $B_0$, the transition energies of the NV probe, $\omega_\mathrm{NV} =
2\pi D\pm\gamma B_0$ (where $\gamma=1.76\times10^{11}\,\mathrm{rad\,s^{-1}T^{-1}}$ is the electron gyromagnetic ratio), can be brought into resonance with environmental spin energies, $\omega_\mathrm{E}$, via the Zeeman effect (Fig.\,\ref{RelaxationFilters}), and the corresponding change in $T_1$ may be measured. We establish a general and robust method to determine the environmental spectral distribution $S(\omega_\mathrm{E})$ by measuring the relaxation time of the NV probe as a function of the external field strength, $T_1(B_0)$. Our method is tested experimentally by measuring the ESR spectrum of the substitutional nitrogen (P1 centres) donor electron spins in type-1b diamond over the range 0-200 MHz. Comparison with theoretical expectations for this known system confirms the validity of this general approach for measuring the ESR spectrum using the NV centre spin-probe in a range of nanoscale applications.

The nitrogen-vacancy centre (see Refs.\,\onlinecite{Doh13} and \onlinecite{Sch14} for extensive reviews) is a point defect in a diamond lattice
comprised of a substitutional atomic nitrogen impurity and an adjacent
 vacancy (Fig.\,\ref{RelaxationFilters}\,(a)). The energy level scheme of the C$_{3v}$-symmetric NV system
(Fig.\,\ref{RelaxationFilters}\,(b)) consists of ground ($^3$A$_2$), excited ($^3$E) and
singlet electronic states. The ground state spin-1 manifold has three
spin sub-levels $\left(\bigl|\,0\bigr\rangle,\bigl|\pm1\bigr\rangle\right)$,
which at zero field are split by 2.87\,GHz. Since the magnetic sublevels have different fluorescence intensities when illuminated with laser light, we can achieve spin-state readout optically\cite{Jel02,Jel06}. The degeneracy
between the $\bigl|\pm1\bigr\rangle$ states may be lifted with the
application of an external field, with a corresponding separation of
2$\times$17.6$\times10^6$\,rad\,s$^{-1}$\,G$^{-1}$, or 5.60\,MHz\,G$^{-1}$, permitting all three states to be accessible via
microwave control, however the $\bigl|\pm1\bigr\rangle$ states are not
directly distinguishable from one another via optical means. By isolating
either the $\bigl|\,0\bigr\rangle\leftrightarrow\bigl|+1\bigr\rangle$ or
$\bigl|\,0\bigr\rangle\leftrightarrow\bigl|-1\bigr\rangle$ transitions, the NV spin system constitutes a controllable, addressable spin qubit.

The large zero-field splitting is fortuitous in the present context in that transitions in the ground-state spin triplet manifold are far detuned from the $\sim$MHz couplings to nitrogen donor electron and $^{13}$C nuclear paramagnetic impurities within the crystal, leaving NV transitions unable to be excited unless brought into resonance using an axial magnetic field ($B_0\approx512$\,G, for an electron spin environment). The weak spin-orbit coupling to crystal phonons (and low phonon occupancy, owing to the large Debye temperature of diamond) leads to longitudinal relaxation of the spin state on timescales of roughly $T_1\sim1-10$\,ms at room temperature. On the other hand, the transverse relaxation time ($T_2$) of the NV spin is the result of dipole-dipole coupling to other spin impurities in the diamond crystal, which occur on timescales of 0.1-1\,$\mu$s in type-1b diamond. Sensing based on the much longer relaxation times can thus be significantly more sensitive than those based on $T_2$. Shot-noise statistics dictate that the minimum magnetic field detectable over a total interrogation time $T = N\tau$, comprised of $N$ cycles of dark time $\tau$, is proportional to $b_\mathrm{min}\propto 1/\sqrt{T\tau}$. As the dark time, $\tau$, is limited by $T_1$ in relaxation based protocols, the latter offer an improvement in sensitivity by a factor of $\sqrt{T_1/T_2}$ over the former for a fixed interrogation time, $T$. Hence, the relatively long relaxation time of the NV ground state and inherent sensitivity to GHz frequencies, together with room temperature operation and optical readout, make it an ideal system for the ESR spectral mapping protocol discussed in this work.

\section{Results}
\subsection{Relaxation of an NV spin in an arbitrary magnetic environment}
At zero field, both NV transitions, ${\bigl|\,0\bigr\rangle\leftrightarrow\bigl|+1\bigr\rangle}$ and
${\bigl|\,0\bigr\rangle\leftrightarrow\bigl|-1\bigr\rangle}$, occur at approximately 2.87\,GHz. These transitions are then split symmetrically by the presence of an axial magnetic field and any axial NV-environment couplings, and move to higher and lower frequencies, respectively, with the resulting shift being directly proportional to the strength of the field (Fig.\,\ref{RelaxationFilters}(c)).
When the transition frequencies of a spin-1 NV system are brought into resonance with those of a particular environmental transition, the two will be able to exchange energy via their magnetic dipole interaction. As the shifts induced by the environment, typically of order MHz, are constantly fluctuating, they act to broaden the NV's transition frequencies, as characterised by their resulting inhomogeneous linewidth, or transverse spin relaxation rate, $\Gamma_2$ (which may be either $\Gamma_2=1/T_2^*$ or $1/T_2$ depending on the whether an ensemble of NVs is used, or just a single NV spin). This results in a resonantly enhanced response of the NV spin relaxation rates ($\Gamma^\pm$) to environmental fields fluctuating at (or within $\Gamma_2$ of) the NV frequency (Fig.\,\ref{RelaxationFilters}(d)).

\begin{figure}
  \centering
  \includegraphics[width=\columnwidth]{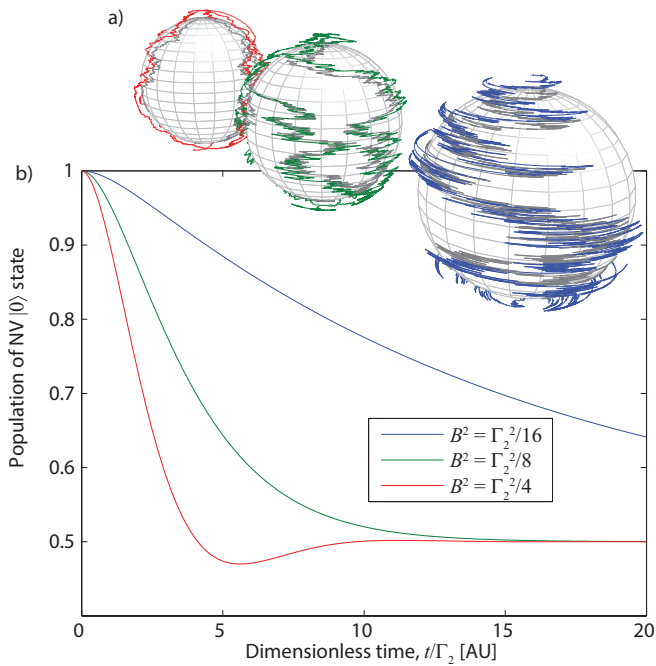}\\
  \caption{ Bloch vector behaviour at constant dephasing rate ($\Gamma_2$), for cases of under-damped ($B>\Gamma_2/2\sqrt2$, red), critically damped ($B=\Gamma_2/2\sqrt2$, green), and over-damped ($B<\Gamma_2/2\sqrt2$, blue) NV-Environment couplings, with the $+z$ axis representing the $|0\rangle$ state of the NV spin, and the $-z$ axis representing the $|-1\rangle$ state. (a) Example realisations of Bloch sphere trajectories. (b) Average Bloch vector trajectories. (c) Population of the NV $|0\rangle$ state of the NV spin ($P_0(t) = \frac{1}{2} + \frac{1}{2}z(t)$). In the under-damped case the Bloch vector is able to rotate appreciably before its transverse projection decays. In the over-damped case the transverse projection of the Bloch vector is always pulled back to the $z$ axis before it can appreciably rotate.  Most practical cases reside in the latter regime, due to the intrinsic dephasing arising from strong coupling of the NV to paramagnetic defects in the diamond crystal.
  }\label{BlochMaps}
\end{figure}

To compute the response of the NV spin to such an environment, we must account for two dominant processes: energy exchange between the NV and the environment (effective coupling rate $B$), which changes the population of the magnetic sublevels of the NV ground state; and the destruction of the phase coherence between these sublevels (dephasing, occurring at a rate of $\Gamma_2$). Depending on the relative strengths of these processes (that is, how many energy exchanges may occur before the NV spin is dephased), the NV spin can exhibit diverse behaviour (Fig.\,\ref{BlochMaps}). However, as there are typically more sources of dephasing than energy exchange (with the latter effect further decreasing when the NV spin and environment are away from resonance), we have $\Gamma_2\gg B$ for the cases considered in this work.

Under this regime, if the NV is initially polarised in its $|0\rangle$ state, and only the $|0\rangle\leftrightarrow|-1\rangle$ transition is being excited, the subsequent population at time $t$ when coupled to a spin system of transition energy $\omega_\mathrm{E}$ is given by (see Supplementary Material for details)
\begin{eqnarray}
  P_0(\omega_\mathrm{E},B_0,B,t)&=&\frac{1}{2}+\frac{1}{2}\exp\left(-2B^2 t\frac{\Gamma_2}{\delta ^2(B_0)+\Gamma_2^2}\right),\,\,\,\,\,\,\,\,\label{MainDecay}
\end{eqnarray}where $\delta = \omega_\mathrm{NV}-\omega_\mathrm{E}$ is the difference in transition frequencies between the NV ($\omega_\mathrm{NV}$) and the environment ($\omega_\mathrm{E}$).

Typical electron spin environments will exhibit a distribution of coupling strengths ($B$) and frequencies ($\omega_\mathrm{E}$), denoted $P_B(B)$ and $S(\omega_\mathrm{E},B_0)$ of roughly the MHz-GHz regime\cite{McG13,Kau13,Hal14}, which must be taken into account to determine the full response of an NV ensemble. This response is given by
\begin{eqnarray}
   \bigl\langle P_0(B_0,t) \bigr\rangle &=&  \int  G(\omega_\mathrm{E},B_0,t)\,S(\omega_\mathrm{E},B_0)\,\mathrm{d}\omega_\mathrm{E},\nonumber\label{EnsembleDecay}
\end{eqnarray}where we identify the function
\begin{eqnarray}
  G(\omega_\mathrm{E},B_0,t) &=&  \int  P_0(\omega_\mathrm{E},B_0,B,t)  P_B(B)\,\mathrm{d}B,\label{FilterDef}
\end{eqnarray}as an environmental noise filter with a Lorentzian point-spread function centred on the NV transition frequency (Eqn.\,\ref{MainDecay}). The NV-relaxation filter, $G$, is tunable via the Zeeman interaction, meaning that it may be directly tuned to specific parts of the environment's spectral distribution, $S(\omega_\mathrm{E},B_0)$, by choosing the strength of the external magnetic field, $B_0$.

If the environment is comprised of some distribution of single electrons, their transition frequencies, $\omega_\mathrm{E}$ (including spectral features such as hyperfine interactions, but centred about the $\gamma B_0$ Zeeman shift), may be brought into resonance with those of the ${\bigl|\,0\bigr\rangle\leftrightarrow\bigl|-1\bigr\rangle}$ transition of the NV (i.e., $\omega_-\sim 2\pi D-\gamma B_0$) by choosing a magnetic field strength $B_0$ such that $B_0 = \pi D/\gamma\approx512\,$G. In other cases, the environment may contain spin-1 or greater systems possessing their own zero-field splitting (See Ref.\,\onlinecite{Kau13} for the case of spin-$\frac{5}{2}$ Gd spins coupled to individual NV centres). If the environmental zero-field splittings are greater than that of the NV, the ${\bigl|\,0\bigr\rangle\leftrightarrow\bigl|+1\bigr\rangle}$ transition of the NV will need to be utilised to ensure that the respective energies are brought into resonance.

\begin{figure*}
  \includegraphics[width=12cm]{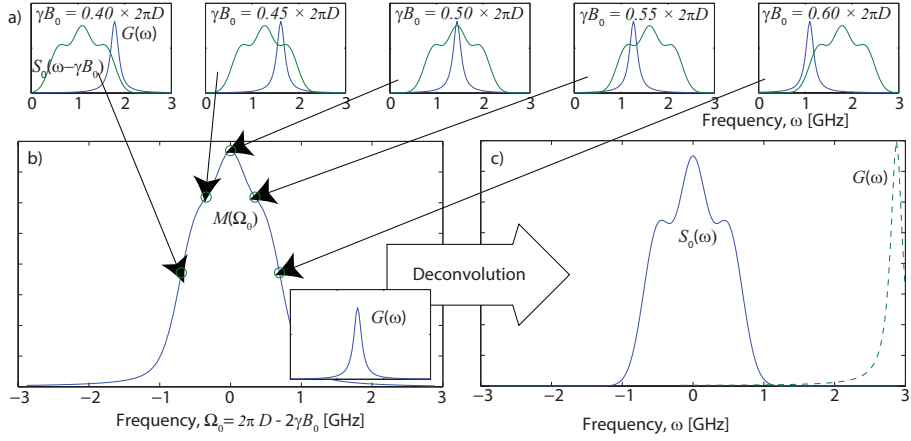}\\
  \caption[Sequence\,(a) By controlling the external field strength the NV filter
  function may be tuned to filter specific regions of the spectral density. (b) The resulting
  measured signal is the convolution of the NV filter function and
  the spectral density of the environment. (c) Given that is known, the spectral density
  may be reconstructed by deconvolution.]{Sequence\,(a) By controlling the external field strength, $B_0$, the NV filter function,
  $G$ may be tuned to filter specific regions of the spectral density, $S$. (b) The resulting
  measured signal, $M(\Omega_0)$, is the convolution of the NV filter function, $G$, and
  the spectral density of the environment, $S$. (c) Given that $G$ is known, the spectral density, $S$,
  may be reconstructed by deconvolving $S$ and $G$ from $M$. }\label{RelaxationScanSchematic}
\end{figure*}
In the following, we discuss how we may take advantage of the relaxation filter to reconstruct the spectral density, $S(\omega_\mathrm{E},B_0)$ of an arbitrary environment.

\subsection{Reconstruction of the environmental spectral
density}
As noted above, the region of the
spectral density sampled by the NV filter functions may be tuned by
controlling the strength of the static external field. This suggests that,
by sweeping the filter function across the entire spectrum, we should in principle be able
to reconstruct it by measuring the relaxation rate of the NV spin for an appropriate range of external field strengths.

 We denote an arbitrary given noise spectrum at zero
field by $S_0(\omega_\mathrm{E})\equiv S(\omega_\mathrm{E},0)$. The distribution at some finite external field
$B_0$ is then $S_0(\omega_\mathrm{E}-\gamma B_0)$. For most cases of practical interest, we assume the shape
of the distribution does not change with $\gamma B_0$ (see Supplementary Materials for more details), although this case can be handled by extension.  Furthermore, we also assume that one of the NV
transitions is sufficiently detuned that it is not sensitive to the
environment, making the overlap with the spectrum insignificant. Even if this is not true, the detuned filter function will
translate with $\gamma B_0$ at the same rate as $S_0(\omega_\mathrm{E}-\gamma B_0)$ (Fig.\,\ref{RelaxationFilters}(d)), and
thus produce a constant shift in the overall measurement that does not change
with $B_0$, which may be later subtracted.

In principle, the filter function, $G$ is fully defined by the hyperfine shift and free induction decay (FID) time ($T_2^*$) of the NV spin (see Supplementary Material for details), which may be characterised by performing a $\pi/2-\tau-\pi/2$ FID (or Ramsey) measurement. The full $G(\omega_\mathrm{E})$ function may also be mapped to arbitrary precision by measuring the response of the NV longitudinal relaxation rate to an applied oscillatory magnetic field of frequency $\omega_\mathrm{E}$, provided the microwave field strength, $B$ (playing the same role as the effective coupling strength of the NV spin to a proximate spin system), is such that $B\ll1/T_2^*$ (to avoid microwave broadening of $G$).

\begin{figure*}
  \includegraphics[width=\textwidth]{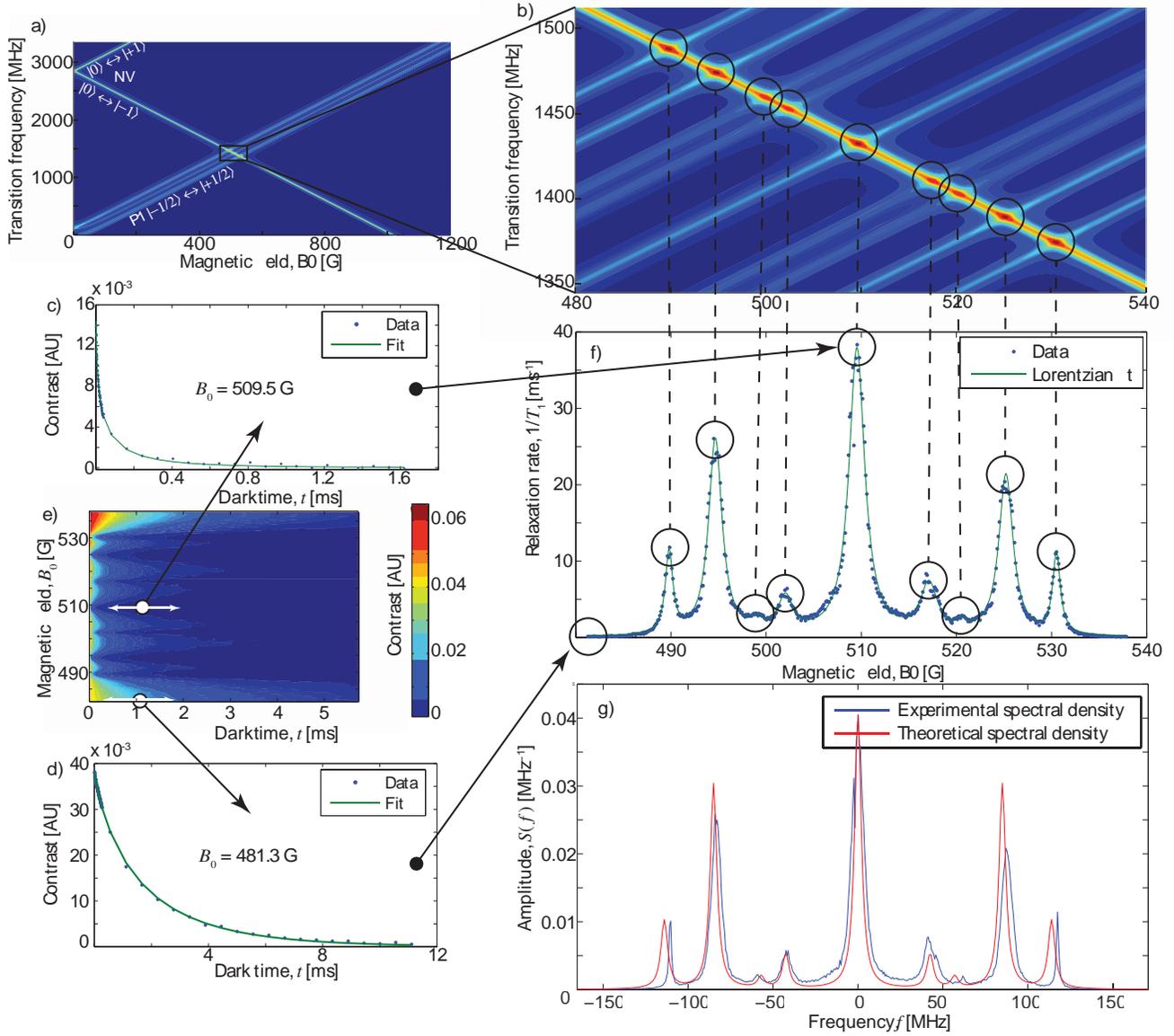}\\
  \caption{(a) Theoretical plot showing the transition frequencies of the NV centre and P1 electron spins vs the strength of an external field aligned along the $\langle111\rangle$ axis. Energy exchange between an NV spin and a nearby P1 spin is achieved when two transition frequencies approach resonance. Broadening of these lines is caused by interactions with other P1 spins and spin impurities within the diamond crystal. (b) Zoomed plot of that in (a), highlighting the hyperfine structure and the corresponding spin conserving and non-spin conserving transitions of the P1 centre. c) \& d) Plots of NV centre contrast vs dark time at 509.5\,G and 481.3\,G respectively. In the case of the former, NV and P1 electron energies are on resonance, resulting in a comparatively rapid decay. In the case of the later, NV and P1 transition energies are too far detuned to facilitate a dipole-mediated resonant energy exchange, meaning that the depolarisation of the NV spin ensemble is dominated by interactions with crystal phonons. e) Measurements of curves such as those shown in c) and d) for 500 magnetic field strengths between 480\,G and 540\,G.
    f) Plot of the external field-dependent relaxation rate, $\Gamma_1(B_0)$, extracted from the data shown in e). Correspondences of the observed features with the P1 spin transitions in b) are indicated explicitly.
    g) Application of the deconvolution procedure to the data in f) yields the spectral density of the spin bath environment surrounding the NV centres in the diamond sample (blue), showing good agreement with theoretical predictions (red - see Supplementary Material for details).\\
         }\label{ExperimentalData}
\end{figure*}

The measured response of the NV relaxation rate, $M(\omega_0,t)$, to $S_0$ for some external field strength, $B_0$ ($\omega_0\equiv\gamma B_0$), is then given by (for brevity, we put $M(\omega_0,t)\equiv \bigl\langle  P_0(t,B_0) \bigr\rangle_\mathrm{meas}$)
\begin{eqnarray}
M(\omega_0,t)&=&\int_{-\infty}^\infty S_0(\omega_\mathrm{E}-\omega_0)G(\omega_\mathrm{E}-2\pi D+\omega_0,t)\,\mathrm{d}\omega_\mathrm{E}.\nonumber\\
\end{eqnarray}By introducing the frequency-space variable $\Omega = \omega_\mathrm{E}-\omega_0$,
and the parameter, $\Omega_0 = 2\pi D-2\omega_0$,
and making use of the symmetry properties of the Lorentzian function, we may write
this integral as a Fourier-space convolution,
\begin{eqnarray}
M(\Omega_0,t)&=&\int_{-\infty}^\infty S_0(\Omega)G(\Omega_0-\Omega,t)\,\mathrm{d}\Omega\nonumber\\
&=& \left(S_0*G\right)_t(\Omega_0).\label{Convolution}
\end{eqnarray}
Given that the filter function is known, the spectral density may thus be reconstructed using an appropriate deconvolution algorithm (see Supplementary Material for details).

In the following section, we provide an experimental example of this protocol applied to the P1 electron spin bath of type-1b diamond.

\subsection{Experimental demonstration}
In order to demonstrate the reconstruction of an environmental spectrum using this technique, we measured the relaxation rate of NV centres subject to a bath of substitutional nitrogen donor (or P1 centre) electron spins. The sample chosen was a synthetic diamond grown with high-pressure high-temperature (HPHT) method, with $\sim$50\,ppm substitutional nitrogen and 1-10\,ppm NV$^-$ concentration. NV centers are oriented along all four symmetry axes, and a solenoid was used to apply an adjustable magnetic field along one of these orientations.

After initializing NV spins in the $\left|0\right\rangle$ state via optical pumping (or the $\left|-1\right\rangle$ state with an additional microwave $\pi$ pulse) we measured the fluorescence following a variable amount of `dark time' (for which the pump laser was off). NVs not aligned with the external field and other defects contribute to the diamond fluorescence, which may decay at rates different from that of the aligned NVs. In order to remove this effect, we employed a form of common-mode rejection (described in Ref.\,\onlinecite{Jar12}). After initializing to either the $\left|0\right\rangle$ or $\left|-1\right\rangle$ state, we measured the fluorescence contrast after each dark time, and subtracted the results for each initial state. The resulting decay of the fluorescence difference indicates the loss of NV population from the initial state, and the magnetic noise these populations are sensitive to. We note here that for cases in which the application of microwaves may be impractical, this technique may be performed in an all optical manner, although this approach will also require the characterisation of other sources of fluorescence within the sample.

The P1 centre, the target of our demonstration, is a substitutional defect of a nitrogen atom in place of a carbon atom in a diamond lattice. The single unbonded electron able to reside along any one of the four crystallographic bond axes, giving rise to four possible hyperfine coupling configurations with the nitrogen nucleus. For field strengths above $\sim100$\,G, the quantisation axis of the P1 electron spin is set by its Zeeman interaction, effectively reducing the number of possible P1 species from four to two. Hence, there exists a 25\% chance that the delocalisation axis of the P1 centre is aligned with the NV axis, giving an axial hyperfine coupling of 114\,MHz, and a 75\% chance that the delocalisation axis is $\arccos\left(-\frac{1}{3}\right)\approx109^o$ to the NV axis, producing an axial hyperfine coupling os 86\,MHz (the spin properties of this defect are discussed in detail in the Supplementary Material). Figures.\,\ref{ExperimentalData}\,(a)\,\&\,(b) show the overlap of the transition frequencies associated with both the NV spin and the P1 centre.

Measurements of the longitudinal spin relaxation of the NV ensemble, $\bigl\langle  P_0(t,B_0) \bigr\rangle_\mathrm{meas}$, were taken at 500 different external magnetic field strengths between 480\,G and 540\,G (Figs.\,\ref{ExperimentalData}\,(c)\,to\,(e)). Before we deconvolve the data set to determine the spectral density, we extract and inspect the magnetic field-dependent component of the NV relaxation rate due to spin-spin relaxation with the environment. Accordingly, the data was fitted using the function given by
\begin{eqnarray}
  P^\mathrm{fit}_0(B_0,t) &=& \exp\left(-\sqrt{\Gamma_1(B_0)t}-Rt\right),
\end{eqnarray}where $R\approx360$\,Hz is the component of the relaxation due to environmental phonons (see Supplementary Material for the derivation of this fitting form). The resulting spin-spin relaxation rates are plotted in Fig.\,\ref{ExperimentalData}\,(f).
As expected, the measurements show that the NV relaxation rate increases when $D-2\omega_0/2\pi = 0,\,\pm86\,\mathrm{MHZ},\,\pm114\,\mathrm{MHz}$, or $B_0 = 490,\,495,\,510,\,525,$ and $530$\,G, corresponding to the conditions under which the NV spin may exchange magnetisation with P1 electron spins. Other features are evident at $B_0=499,\,502,\,517\,$ and 520\,G, which correspond to processes that change the magnetisation of the NV spin and the P1 nuclear spin, but leave the P1 electron spin unchanged (as seen previously in Ref.\,\onlinecite{Arm10}). These transitions are not energy conserving and are thus comparatively weak, leaving them partially obscured by the large spin-phonon relaxation effect of the NV spin, resulting from two-phonon Orbach\cite{Red91} and two-phonon\cite{Wal68} Raman processes.

To obtain the spectral distribution, we carry out the deconvolution over the entire set of data, $\bigl\langle  P_0(t,B_0) \bigr\rangle_\mathrm{meas}$, and plot the spectral distribution in frequency space, $S(f)$, in Fig.\,\ref{ExperimentalData}\,(g), demonstrating good agreement with the corresponding theoretical expectations (see Supplementary Material for details). The deconvolution to the spectral density is shown to remove much of the broadening seen in the raw relaxation data in Fig.\,\ref{ExperimentalData}\,(f), and thus provides a better measure of the environmental dynamics. Some small discrepancies between the resulting spectrum and the theoretical result are evident, however, we note that whilst the theory incorporates effects such as hyperfine couplings and dephasing rates, more complicated effects such as g-factor anisotropy and enhancement, interactions with other paramagnetic impurities, and other strain-related phenomena, have not been included. Such effects are highly sample-dependent and thus difficult to predict in general terms; however the technique developed in this work provides an ideal means by which to facilitate their investigation. Finally, we note that although this demonstration involves ensembles of NV and P1 centres, the detection is highly local as the response of each NV is dominated by its nearest P1 centre: for this sample (50\,ppm P1) the mean distance to the nearest P1 centre is about 2.7\,nm\,\cite{Hal14}.

\section{Discussion}
In this work we have presented a general method for extracting the spectral distribution of an arbitrary electronic environment based on tuning a spin-1 NV probe system, via controlled application of an external field, into resonance with the transitions of the target electronic system. The method was tested using an ensemble of NV centres in a type-1b diamond sample to determine the spectral distribution of the P1 spin bath, showing excellent agreement with the theoretical expectations. Our relaxation based ESR method has a number of advantages over existing techniques. Measurements of the NV relaxation in general do not require microwave control, and thus require no manipulation of the sample. With relaxation times much longer than dephasing times, $T_1$ based protocols can be significantly more sensitive to ESR detection. Finally, even in the ensemble case demonstrated here, the NV spin relaxation is dominated by local interactions with the environment affords an effective spatial resolution of a few nanometres. By extending this technique to the single-NV probe case (or other solid-state single spin systems), determination and characterisation of nanoscale ESR spectra of single electronic systems will be possible.

\acknowledgments{The authors gratefully acknowledge discussions with L. McGuinness, and J. Wood. This work was supported in part by the Australian Research Council (ARC) under the Centre of Excellence scheme (project No. CE110001027). LCLH acknowledges the support of an ARC Laureate Fellowship (project no. FL130100119). DB acknowledges support from the the AFOSR/DARPA QuASAR program and the German-Israeli Project Cooperation (DIP) program.
}

\bibliography{Spectrum}

\newpage
\appendix
\begin{widetext}
\section{Modeling of the NV relaxation process}\label{APPRelaxation}
 Here we develop the theory of relaxation based sensing as used in the main text of this work. As we are considering axial magnetic field strengths, $B_0$, such that $B_0\sim D/2 \sim 512\,$G, only the ${|0\rangle\leftrightarrow|-1\rangle}$ transitions of the NV spin will be appreciably excited, meaning we can disregard any population of the $|+1\rangle$ state. The time evolution of the associated density matrix is described by
\begin{eqnarray}
  \frac{\mathrm{d}\rho_T}{\mathrm{d} t} &=& -i\bigl[\mathcal{H}_T(t),\rho_T\bigr],
\end{eqnarray} where $\rho_T$ represents the combined
density matrix of the entire spin + environment system. The full Hamiltonian
is given by $\mathcal{H}_T = \mathcal{H}_0 + \mathcal{V}+\mathcal{H}_E$ where
$\mathcal{H}_0$ and $\mathcal{H}_E$ are the self Hamiltonians of the NV
centre and environment respectively. The coupling of the environment to the
NV is described by the full dipolar interaction due to all spins in the
environment:
\begin{eqnarray}
 \mathcal{V}  &=& \frac{\mu_0}{4\pi}\hbar\gamma_{nv}\gamma_{E} \sum_i\frac{1}{R_i^3}\left[\vec{\mathcal{P}}\cdot\vec{\mathcal{S}}_{i}-3 \frac{1}{R_i^2}\left(\vec{\mathcal{P}}\cdot\mathbf{R}_i\right)\left(\mathbf{R}_i\cdot\vec{\mathcal{S}}_{i}\right)\right],
\end{eqnarray}
which includes both transverse and longitudinal components, proportional to
$\mathcal{P}_{x,y}$ and $\mathcal{P}_z$ of the NV spin respectively. The
latter have a pure dephasing effect, resulting in an additional contribution
to the intrinsic dephasing rate of the NV. As relaxation processes occurr on timescales that are much longer that the typical interaction timescales of the environmental constituents, the resulting dephasing will be purely exponential.
These effects may thus be modeled using a master equation approach for the
reduced density matrix for the ensemble averaged dynamics of the NV, $\rho$,
as follows
\begin{eqnarray}
  \frac{\mathrm{d}\rho}{\mathrm{d} t} &=& -i\bigl[\mathcal{H}(t),\rho\bigr] +\sum_{NV,i}\left( \mathcal{L}_i\rho\mathcal{L}_i^\dag - \frac{1}{2}\left\{\mathcal{L}_i^\dag\mathcal{L}_i,\rho\right\}\right),\,\,\,\,\,\,\,\,\,\,\,\,\,
\end{eqnarray} where, in the present context, $\mathcal{L}_i$ is the Lindbladian
 operator corresponding to a pure dephasing process on spin $i$, and is given
 by $\mathcal{L}_i = \sqrt{2\Gamma_2}\mathcal{S}_{z,i}$. The total dephasing rate due to both the local
 crystal environment and the longitudinal coupling to the environment is
 given by $\Gamma_2 = \left(T_2^*\right)^{-1} + \Gamma_2^\mathrm{nv-E}$. The
 timescale of the intrinsic dephasing process is described using the
 inhomogeneous linewidth, $\left(T_2^*\right)^{-1}$, since the transverse phase
 accumulation occurs in the absence of any pulsed microwave control. Subtle
 tuning effects that modify the sensitivity of this technique to various
 parts of the environmental spectral density may be achieved by changing the
 intrinsic dephasing rate via dynamic decoupling techniques.

In what follows, owing to the strong intra-environment and comparatively weak
NV-environment couplings, we will treat the coupling of the environment to
the transverse components of the NV spin as a semiclassical oscillatory
field (these simplifications will be rigorously justified in what follows),
\begin{eqnarray}
\mathcal{V} &=& \mathcal{B}e^{i\omega t} + \mathcal{B}^\dag e^{-i\omega
t},\end{eqnarray}where $\mathcal{B} = B_x(\omega)\mathcal{S}_x +
B_y(\omega)\mathcal{S}_y$; and $B_x$ and $B_y$ are the $x$ and $y$ components
of the magnetic field. The frequency spectrum is determined by analysing the
interaction between environmental constituents, as described by
$\mathcal{H}_E$. To make the solution tractable, we change to the interaction
picture. The transformed
equation of motion is given by
    \begin{eqnarray}
      \frac{\mathrm{d}}{\mathrm{d} t}\rho_I(t)    &=&  -i\bigl[ \mathcal{V}_I(t),\rho_I(t)\bigr]
      + \Gamma_2\bigl(\mathcal{S}_z\rho_I(t)\mathcal{S}_z -  \rho_I(t)\bigr)\label{Relaxation3by3system},\,\,\,\,\,\,\,\,\,\,\,\,
         \end{eqnarray}with the interaction Hamiltionian given by
         $\mathcal{V}_I = e^{i\mathcal{H}_0t}\mathcal{V}e^{-i\mathcal{H}_0t}$.

We are then interested in determining the rate at which the NV spin relaxes
to its equilibrium state under the influence of the environment. We proceed
by reducing the $3\times3$ system of first order linear differential
equations described by equation\,\ref{Relaxation3by3system} to a higher order
differential equation for $P_0 \equiv\rho_{00}$. We then wish to solve this equation,
together with the initial conditions of $\rho_{ij}=0$ unless $i=j=0$, in
which case we have $\rho_{00} = 1$, representing the initial polarisation of
the NV spin in the $|0\rangle$ state.

To gain insight into the expected analytic solution for the spin-1 NV centre,
we consider the simplified case in which only one of the transitions of the
NV centre is excited by the environment, and the other is assumed to be too
far detuned to have any effect on the population of the spin states. This
simplifies the analysis dramatically, yet demonstrates the main properties of
relaxation based detection. Incidentally, this simplification is still
applicable to a spin-1 system for cases of significantly strong Zeeman
splittings between the $|\pm1\rangle$ states. This ensures that one
transition will be excited by the environment, whilst the other is not. This
forms the basis of the technique by which environmental spectra may be
mapped.

The equation of motion
for $P_0(\omega_\mathrm{E},B_0,B,t)\equiv\rho_{00}(t)$ is
\begin{eqnarray}
   \frac{\mathrm{d}^3P_0}{\mathrm{d}t^3}+ 2\Gamma  \frac{\mathrm{d}^2P_0}{\mathrm{d}t^2}+\left(\Gamma^2+\delta ^2+2B^2\right)\frac{\mathrm{d}P_0}{\mathrm{d}t} + 2\Gamma  B^2 P_0 -\Gamma B^2&=& 0,\,\,\,\,\,\,\,\,\,\,\,\label{ClassicalDE}
\end{eqnarray}where $B \equiv \left\langle\mathcal{B}^2\right\rangle^\frac{1}{2}$ is the second moment of the strength of the effective magnetic field operator, $\mathcal{B}$.

\subsection{Response to a monochromatic transverse field}
\subsubsection{Resonant case}
For the case where the frequency of the environment is resonant with the
transition frequency between the probe's spin states ($\delta = 0$), the solution of Eq.\,\ref{ClassicalDE} is
\begin{eqnarray}
  P_0(\omega_\mathrm{E},B_0,B,t) &=& \frac{1}{2}+\frac{1}{2}\exp\left(-\frac{\Gamma_2 t}{2} \right)\left[\cosh\left( \frac{t}{4}\sqrt{{\Gamma_2^2}-8{B^2}}\,\right)   +    \frac{\Gamma_2}{\sqrt{\Gamma_2^2-8B^2}}\sinh\left(  \frac{t}{4}\sqrt{{\Gamma_2^2}-8{B^2}}\,\,\right) \right].
\end{eqnarray}
Typically the spin based environments in which we are interested
couple weakly to the NV spin as compared with its intrinsic dephasing rate,
implying $\Gamma_2\gg B$. In fact, even a strong coupling will also induce
additional dephasing, so even in a worst case scenario we are guaranteed
$\Gamma_2> B$. In this limit, we have
\begin{eqnarray}
  \left.P_0(B,t)\right|_{\delta=0} &=&  \frac{1}{2}+\frac{1}{2}\exp\left(-2\frac{B^2}{\Gamma_2}t\right).
\end{eqnarray}Hence the resonant (and therefore maximal) longitudinal relaxation rate is given by
\begin{eqnarray}
  \Gamma^\mathrm{max}_1  = 2\frac{B^2}{\Gamma_2} = 2T^*_2\left\langle\mathcal{B}^2\right\rangle\label{RelaxationResonantDecay}.
\end{eqnarray}
\subsubsection{General case}
When a finite detuning, $\delta$, exists, we may examine relative importance of the terms within Eq.\,\ref{ClassicalDE},
subject to rescaling $t$ in terms of the decay time from the resonant solution. That is,
if we consider the dimensionless variable $T = \Gamma^\mathrm{max}_1t$ and retain terms up to and including order $\mathcal{O}\left(\frac{B^2}{\Gamma_2^2}\right)$, the solution for an arbitrary detuning becomes
\begin{eqnarray}
  \left.P_0(B,t)\right|_{\delta=0}&=&\frac{1}{2}+\frac{1}{2}\exp\left(-2B^2 t\frac{\Gamma_2}{\Gamma_2^2+\delta ^2}\right).\label{oneexcitedtransition}
\end{eqnarray}For zero detuning, we recover the previous result (Eq.\,\ref{RelaxationResonantDecay}).
For finite detuning, the relaxation rate is modified by a Lorentzian factor with a FWHM of $\Gamma_2$.
The complete decay profile is then obtained by integrating this expression over the spectral density of the environment, implying that the $\delta$-dependent relaxation rate acts acts to filter out the environmental spectrum about $\delta \sim 0$.

\subsection{Response to a transverse field with an arbitrary spectral density}\label{AppArbSpecDens}

Even without considering the specifics of the spectral density, the response of the NV spin to an arbitrary spin bath can vary remarkably due the geometric proximity and arrangement of the bath relative to the NV centre. The definitions of the NV spin relaxation and the corresponding filter function are given by
\begin{eqnarray}
   M(B_0,t)  &=&  \int  G(\omega_\mathrm{E},\omega_0,t)\,S(\omega_\mathrm{E},\omega_0)\,\mathrm{d}\omega_\mathrm{E},\label{AppDefOfM}
\end{eqnarray},and
\begin{eqnarray}
  G(\omega_\mathrm{E},\omega_0,t) &=&  \int  P_0(\omega_\mathrm{E},\omega_0,B,t)  P_B(B)\,\mathrm{d}B,\label{AppDefOfG}
\end{eqnarray} respectively, where the filter function, $G(\omega_\mathrm{E},\omega_0)$ acts to filter out regions of the spectral density (as dependent on the external field strength, $\omega_0$), and depends explicitly on the geometric arrangement of the environmental constituents. Ultimately, given some measurement record and filter function, it is expression \ref{AppDefOfM} that must be deconvolved to reproduce the spectral density, $S(\omega_\mathrm{E})$.
In this section, we consider the effects of the geometric arrangement of the environment on the filter function, $G$, for a general spectral density; and specific cases of important spectral densities, namely that due to the internal P1 nitrogen donor electron spin bath in type-1b diamond and its corresponding surface spin bath, are considered below in Section\,\ref{APPP1centre}.

\subsubsection{Response to an internal (bulk) spin bath}\label{AppBulk}
We note that the coupling of the NV to a bath spin located at some distance $r$ may be written $B\equiv b(\theta,\phi)/r^3$, where the specific details of the angular dependence of the coupling are incorporated into the parameter $b(\theta,\phi)$. We note that is is necessary to omit further discussion of $b$ until Section\,\ref{APPP1centre}, as different environmental processes will be more readily detectable by the NV spin at different relative angles.
Unlike the transverse (spin echo) case, where the precession of the NV spin vector in the $x-y$ plane is sensitive to \emph{all} longitudinal field sources, the effect on the longitudinal projection is dominated by the coupling to the nearest P1 electron spin. As such, we may write $P_B(B)=P_r(r)P_{\theta,\phi}(\theta,\phi)$, where $r,\theta,\phi$ are the spherical coordinates associated with the distribution of field sources.


From Ref.\,\onlinecite{Hal14}, the distribution of distances from a given NV centre to its nearest spin impurity is given by
\begin{eqnarray}
  \mathrm{P}_r(r) &=& 4\pi n r^2\exp\left(\frac{4}{3}\pi n r^3\right),
\end{eqnarray}where $n$ is the average density of impurities in the bath. Substituting this expression into Eq.\,\ref{AppDefOfM}, we find
\begin{eqnarray}
 M(\omega_0,t)  &=& \frac{1}{2} +\frac{1}{3} \sqrt{2\pi } b(\theta,\phi) n \,\int \sqrt{\frac{\Gamma _2 t}{\Gamma _2^2+\delta ^2}} \,\mathfrak{\mathfrak{G}}_{0,3}^{3,0}\left(\left.\frac{8 b^2(\theta,\phi) n^2 \pi ^2 t \Gamma _2}{9 \left(\Gamma _2^2+\delta ^2\right)}\right|
\begin{array}{c}
 -\frac{1}{2},0,\frac{1}{2} \\
\end{array}
\right)\,P_{\theta,\phi}(\theta,\phi)\,\mathrm{d}\theta\,\mathrm{d}\phi\,S(\omega_\mathrm{E},\omega_0)\,\mathrm{d}\omega_\mathrm{E} \nonumber\\
&\sim& \int\exp\left(-\frac{2\pi}{3} \left(2\pi  n^2 {t} \right)^{1/2}\bigl\langle |b|\bigr\rangle\,\sqrt{\frac{\Gamma _2}{\Gamma _2^2+\delta ^2}}\,\right)\,S(\omega_\mathrm{E},\omega_0)\,\mathrm{d}\omega_\mathrm{E},
\end{eqnarray}for smal $t$, where $\mathfrak{G}$ is the Meijer-G function, and $\bigl\langle |b|\bigr\rangle=\int|b(\theta,\phi)|\,P_{\theta,\phi}(\theta,\phi)\,\mathrm{d}\theta\,\mathrm{d}\phi$. Thus, we identify the filter function associated with environments inside the diamond lattice to be
\begin{eqnarray}
  G_\mathrm{in}(\omega_\mathrm{E},\omega_0) &=&  A_\mathrm{in}\sqrt{\frac{\Gamma _2}{(\omega_\mathrm{E} - D + \omega_0) ^2+\Gamma _2^2}},
\end{eqnarray}where $A_\mathrm{in}$ is a constant, associated with the geometry of the bath, that may be renormalised.

\subsubsection{Response to an external (surface) spin bath}
In contrast to the bulk spin bath case, spins on the surface are unable to exist arbitrarily close to the NV centre. Typically we consider samples in which NV centres exist at some depth $h+\delta h$ below the surface, with $h$ being the mean depth, and $\delta h$ a normally distributed variable with variance $\left\langle\delta h^2\right\rangle\ll h^2$. In this case, an individual NV spin is exposed to many bath spins, meaning that the effective coupling distribution, $\mathrm{P}_B(B)$, is normally distributed.

In this case, we may expand Eq.\,\ref{oneexcitedtransition} for small $t$, which, upon substitution into Eq.\,\ref{AppDefOfM} and averaging over $\mathrm{P}_B(B)$ gives
\begin{eqnarray}
\bigl\langle P_0(t)\bigr\rangle &\sim& 1-4t\left\langle B^2\right\rangle\left(\int\frac{\Gamma _2}{\Gamma _2^2+\delta ^2}\,S(\omega_\mathrm{E},\omega_0)\,\mathrm{d}\omega_\mathrm{E}\right)
\end{eqnarray}

 Thus, we identify the filter function associated with environments outside the diamond lattice to be
\begin{eqnarray}
  G_\mathrm{out}(\omega_\mathrm{E},\omega_0) &=&   A_\mathrm{out}\frac{\Gamma_2}{\left(\omega_\mathrm{E} - D+\omega_0\right) ^2+\Gamma_2^2}.
\end{eqnarray}

\section{Derivation of the deconvolution algorithm}\label{APPDevonvDerivation}
Although we wish to invert Eq.\,\ref{AppDefOfM} for the spectrum $S_0 \equiv S(\omega_\mathrm{E},0)$, we must also deal with the intrinsic shot noise, $\eta(\Omega_0)$ that arises during the measurement process. Thus, we are interested in solving
\begin{eqnarray}
M(\Omega_0)&=& \left(S_0*G\right)(\Omega_0) + \eta(\Omega_0),\label{ConvolutionWithNoise}
\end{eqnarray}where the effective spectral features of $\eta$ may be inferred from the fact that the errors associated with any two measured time points are uncorrelated from one-another. Our goal is thus to find some function, $H$, such that a least-squares estimate of $S_0$ is given by
\begin{eqnarray}
  \overline{{S}}_0 &=& H*M.\label{deconvolve}
\end{eqnarray} In order to deal with products of the relevant functions, rather than convolutions, we switch to Fourier space via the Fourier transform, denoted $\mathfrak{F}$.

The mean-square error in $\bar{S}_0$ is given by
\begin{eqnarray}
  \mathrm{LSE} &=& \mathrm{Ex}\left\{\left|\mathfrak{F}\left(\bar{S}_0\right) -\mathfrak{F}\left(S_0\right)\right|^2\right\} \nonumber\\
  &=& \mathrm{Ex}\left\{\left|\mathfrak{F}\left(H\right)\mathfrak{F}\left(\Gamma_1\right) -\mathfrak{F}\left(S_0\right)\right|^2\right\} \nonumber\\
  &=& \mathrm{Ex}\left\{\left|\mathfrak{F}\left(H\right)\left[\mathfrak{F}\left(S_0\right)\mathfrak{F}\left(G\right) + \mathfrak{F}\left(\eta\right)\right] -\mathfrak{F}\left(S_0\right)\right|^2\right\} \nonumber\\
   &=& \mathrm{Ex}\left\{\bigr(\mathfrak{F}\left(S_0\right)\left[\mathfrak{F}\left(H\right)\mathfrak{F}\left(G\right)-1\right]\bigr)\bigl(\mathfrak{F}\left(S_0\right)\left[\mathfrak{F}\left(H\right)\mathfrak{F}\left(G\right)-1\right] \bigr)^*\right\}+ \mathrm{Ex}\left\{\bigr(\mathfrak{F}\left(H\right) \mathfrak{F}\left(\eta\right)\bigr)\bigl( \mathfrak{F}\left(H\right) \mathfrak{F}\left(\eta\right)\bigr)^*\right\} \nonumber\\
      &&+ \mathrm{Ex}\left\{\bigr(\mathfrak{F}\left(S_0\right)\left[\mathfrak{F}\left(H\right)\mathfrak{F}\left(G\right)-1\right]\bigr)\bigl(\mathfrak{F}\left(H\right) \mathfrak{F}\left(\eta\right)\bigr)^*\right\} + \mathrm{Ex}\left\{\bigr(\mathfrak{F}\left(H\right) \mathfrak{F}\left(\eta\right)\bigr)\bigl(\mathfrak{F}\left(S_0\right)\left[\mathfrak{F}\left(H\right)\mathfrak{F}\left(G\right)-1\right]\bigr)^*\right\} \nonumber\\
   &=& \left[\mathfrak{F}\left(H\right)\mathfrak{F}\left(G\right)-1\right]\left[\mathfrak{F}\left(H\right)\mathfrak{F}\left(G\right)-1\right] ^*\mathrm{Ex}\left\{\left|\mathfrak{F}\left(S_0\right)\right|^2 \right\}
   + \mathfrak{F}\left(H\right) \mathfrak{F}^*\left(H\right)\mathrm{Ex}\left\{\left|\mathfrak{F}\left(\eta\right)\right|^2\right\},
           \end{eqnarray}where the last line follows from the fact that the noise and the environmental spectrum are uncorrelated. Minimising the least-squares estimate of this quantity with respect to $\mathfrak{F}\left(H\right)$ (see Section\,\ref{APPDevonvDerivation}), we find
\begin{eqnarray}
  H &=& \frac{\mathfrak{F}^*(G)\left|\mathfrak{F}(S_0)\right|^2}{\left|\mathfrak{F}(G)\right|^2\left|\mathfrak{F}(S_0)\right|^2+\left|\mathfrak{F}(\eta)\right|^2}.
\end{eqnarray}Thus, we have
\begin{eqnarray}
  \overline{S}_0 &=& \mathfrak{F}^{-1}\bigl(\mathfrak{F}(H)\mathfrak{F}(M)\bigr).
\end{eqnarray}This shows that determination of the appropriate deconvolution filter requires knowledge of $S_0$, which is what we are actually trying to measure, however, since the measurement record, $M(\omega_0)$ contains many spectral features similar to $S_0$, $M(\omega_0)$ provides an ideal starting point with which to implement an iterative procedure using this method.

\section{Theoretical description of the coupled NV-P1 system}\label{APPP1centre}
In this section, we discuss the features we expect to be evident in the P1 centre spectrum by examining the effect of a P1 centre on the magnetic field dependent relaxation rate of a near-by NV centre. We conclude this section by demonstrating the equivalence of the semi-classical approach used in this work, and a quantum mechanical treatment of the NV-P1 interaction.

\subsection{P1 Hamiltonian}
 The Hamiltonian of a P1 centre is given by
\begin{eqnarray}
  \mathcal{H}_\mathrm{P1} &=& \vec{\mathcal{S}}\cdot\mathbf{A}_\mathrm{P1}\cdot\vec{\mathcal{I}} + \mathbf{B}_0\cdot\left(\gamma_\mathrm{e}\vec{\mathcal{S}}+\gamma_\mathrm{N}\vec{\mathcal{I}}\right) + \vec{\mathcal{I}}\cdot\mathbf{Q}_\mathrm{P1}\cdot\vec{\mathcal{I}},\,\,\,\,
\end{eqnarray}where $\mathbf{A}_\mathrm{P1}$ is the hyperfine tensor describing the coupling between the P1 electron, $\vec{\mathcal{S}}$, and $^{14}$N nuclear spin, $\vec{\mathcal{I}}$; and $\mathbf{Q}$ is the quadrupole splitting of the nuclear spin. For all field strengths at which there is an appreciable overlap of this spectrum with the NV spin filter function, $B_0\sim512\,$G, we find that the eigenstates of the P1 centre electron spin are predominantly dictated by the external magnetic field. In this instance, the Hamiltonian of the P1 centre becomes
 \begin{eqnarray}
  \mathcal{H}^\mathrm{on}_\mathrm{P1} &=& A_{z}\mathcal{S}_z\mathcal{I}_z + A_{x}\left(\mathcal{S}_x\mathcal{I}_x +\mathcal{S}_y\mathcal{I}_y \right) + B_0\left(\gamma_\mathrm{e}\mathcal{S}_z+\gamma_\mathrm{N}\mathcal{I}_z\right) + Q_\mathrm{P1}\mathcal{I}_z^2,
\end{eqnarray}for cases where the P1 axis is aligned with the external magnetic field, where $A_{z} = 114$\,MHz, and $A_{x} = 81.3$\,MHz. If the P1 axis is aligned along one of the three other bond directions not aligned with the field, the Hamiltonian may be transformed via the rotation operator $ \mathcal{R} = \exp\left(-i\mathcal{I}_y\theta\right) = 1-i\mathcal{I}_y\sin(\theta)-\mathcal{I}_y^2\left(1-\cos(\theta)\right)$ (The other two axes may be realised via a trivial rotation about the $z$ axis), and is given by
 \begin{eqnarray}
  \mathcal{H}^\mathrm{off}_\mathrm{P1} &=& a_z\mathcal{S}_z\mathcal{I}_z + a_x \left(\mathcal{S}_x\mathcal{I}_x +\mathcal{S}_y\mathcal{I}_y \right) + B_0\left(\gamma_\mathrm{e}\mathcal{S}_z+\gamma_\mathrm{N}\mathcal{I}_z\right) + Q'_\mathrm{P1}\mathcal{I}_z^2,
\end{eqnarray}where $a_z = \frac{1}{9} \left(8 A_x + Az\right)=85$\,MHz, and $a_x=\frac{1}{9} \left(5 A_x + 4Az\right)=99$\,MHz.

\subsection{Coupling of the P1 environment to the NV centre}

The interaction between the NV spin and the P1 nuclear spin is ignored on account of its comparative weakness. For the interaction between the NV and the P1 electron, the magnetic dipole interaction is given by
\begin{eqnarray}
  \mathcal{H}_\mathrm{int} &=& B_\mathrm{int}\left[\vec{\mathcal{P}}\cdot\vec{\mathcal{S}}-3\left(\hat{\mathbf{r}}\cdot\vec{\mathcal{P}}\right)\left(\hat{\mathbf{r}}\cdot\vec{\mathcal{S}}\right)\right],
\end{eqnarray}where $\hat{\mathbf{r}}$ is the unit separation vector between the NV and P1 centres, and
\begin{eqnarray}
  B_\mathrm{int} &=& \frac{\mu_0\hbar\gamma_e^2}{4 \pi r^3},
\end{eqnarray} is the effective dipolar coupling strength.

The components of the NV-P1 interaction responsible for the relaxation of the NV spin are those coupling to its transverse components, namely $\mathcal{S}_x$ and $\mathcal{S}_y$. Without loss of generality, we may rewrite this interaction as an effective quantum mechanical magnetic field, $\vec{\mathcal{B}}$, where $\mathcal{B}_{x,y,z}$ couples to $\mathcal{S}_{x,y,z}$.

To make use of Eq.\ref{AppDefOfM}, we make the semi-classical approximation, and assume that NV-P1 interaction plays very little part in determining the dynamics of the environment. The problem of determining the environmental spectrum then reduces to solving for the environmental evolution exclusively under its own influence as follows.

The effective field strength associated with the allowed NV-P1 $|0,\downarrow\rangle\leftrightarrow|+1,\uparrow\rangle$ transition is given by
\begin{eqnarray}
 B_\mathrm{all} &\equiv& \frac{1}{2}\sqrt{\left(B_{xx}+B_{yy}\right)^2 + \left(2B_{xy}\right)^2}\nonumber\\
 &=& \frac{3}{2}B\sin^2\left(\Theta\right),
 \end{eqnarray}and that due to the disallowed $|0,m_S,m_I\rangle\leftrightarrow|+1,m_S+1,m_I-1\rangle,\,|0,m_S,m_I\rangle\leftrightarrow|+1,m_S-1,m_I+1\rangle$ transitions is given by
 \begin{eqnarray}
 B_\mathrm{dis} &\equiv& \frac{1}{2}\sqrt{B_{xz}^2 + B_{yz}^2}\nonumber\\
 &=&\frac{3}{4 }B\sin\left(2\Theta\right)
 \end{eqnarray}

To determine the effect of the P1 environment, we compute the autocorrelation functions associated with the field components above. Interactions between environmental components may be modeled by damping these autocorrelation functions with a decaying exponential, $\exp\left(-\Gamma_\mathrm{P1}t\right)$ to describe their relaxation due to mutual flip-flop processes with corresponding relaxation rate $\Gamma_\mathrm{P1}$.

The corresponding spectra may then be found by computing the Fourier transforms of the autocorrelation functions. From this, we find the spectra associated with the allowed transitions to be
\begin{eqnarray}
  S^\mathrm{on}_\mathrm{all}(\omega) &=& \frac{1}{6\pi}\left[\frac{\Gamma_{\mathrm{P1}}}{\Gamma^2_{\mathrm{P1}}+(\omega\pm\omega_0)^2} +  \frac{\Gamma_{\mathrm{P1}}}{\Gamma^2_{\mathrm{P1}}+(\omega\pm\omega_0+A_z)^2} +  \frac{\Gamma_{\mathrm{P1}}}{\Gamma^2_{\mathrm{P1}}+(\omega\pm\omega_0-A_z)^2}\right],
\end{eqnarray}
and
\begin{eqnarray}
  S^\mathrm{off}_\mathrm{all}(\omega) &=& \frac{1}{6\pi}\left[\frac{\Gamma_{\mathrm{P1}}}{\Gamma^2_{\mathrm{P1}}+(\omega\pm\omega_0)^2} +  \frac{\Gamma_{\mathrm{P1}}}{\Gamma^2_{\mathrm{P1}}+(\omega\pm\omega_0+a_z)^2} +  \frac{\Gamma_{\mathrm{P1}}}{\Gamma^2_{\mathrm{P1}}+(\omega\pm\omega_0-a_z)^2}\right],
\end{eqnarray}
for cases of on and off axis P1 centres respectively. Taking the relative proportions of on and off-axis P1 centres to be 25\% and 75\% respectively, we find the overall spectrum associated with the allowed transitions to be
\begin{eqnarray}
  S_\mathrm{all}(\omega) &=& \frac{1}{4} S^\mathrm{on}_\mathrm{all}(\omega)+\frac{3}{4}S^\mathrm{off}_\mathrm{all}(\omega).
\end{eqnarray}

Similarly, the spectra associated with the disallowed transitions are given by
\begin{eqnarray}
  S^\mathrm{on}_\mathrm{dis}(\omega) &=& \frac{1}{4\pi}\left[ \frac{\Gamma_{\mathrm{P1}}}{\Gamma^2_{\mathrm{P1}}+(\omega\pm\omega_0+\lambda_1)^2} +  \frac{\Gamma_{\mathrm{P1}}}{\Gamma^2_{\mathrm{P1}}+(\omega\pm\omega_0-\lambda_2)^2}\right],
\end{eqnarray}
and
\begin{eqnarray}
  S^\mathrm{off}_\mathrm{dis}(\omega) &=& \frac{1}{4\pi}\left[ \frac{\Gamma_{\mathrm{P1}}}{\Gamma^2_{\mathrm{P1}}+(\omega\pm\omega_0+\lambda_3)^2} +  \frac{\Gamma_{\mathrm{P1}}}{\Gamma^2_{\mathrm{P1}}+(\omega\pm\omega_0-\lambda_4)^2}\right],
\end{eqnarray}respectively, where
\begin{eqnarray}
  \lambda_1&=&\sqrt{2 A_x^2 +  \left(\omega_0+\frac{A_z}{2}  \right)^2}\sim \omega_0+\frac{A_z}{2}\nonumber\\
  \lambda_2&=&\sqrt{2 A_x^2 +  \left(\omega_0-\frac{A_z}{2}  \right)^2}\sim \omega_0-\frac{A_z}{2}\nonumber\\
  \lambda_3&=&\sqrt{2 a_x^2 +  \left(\omega_0+\frac{a_z}{2}  \right)^2}\sim \omega_0+\frac{a_z}{2}\nonumber\\
  \lambda_4&=&\sqrt{2 a_x^2 +  \left(\omega_0-\frac{a_z}{2}  \right)^2}\sim \omega_0-\frac{a_z}{2}\nonumber.
\end{eqnarray}The spectrum associated with the disallowed transitions is then
\begin{eqnarray}
  S_\mathrm{dis}(\omega) &=& \frac{1}{4} S^\mathrm{on}_\mathrm{dis}(\omega)+\frac{3}{4}S^\mathrm{off}_\mathrm{dis}(\omega).
\end{eqnarray}

By employing the full spectrum, $S(\omega_E)=S_\mathrm{all}(\omega_E)+S_\mathrm{dis}(\omega_E)$, in equation\,\ref{AppDefOfM}, we find the resulting external field-dependent relaxation rate of the NV centre to be\begin{eqnarray}
\Gamma_1(\omega_0) &=& \frac{\left\langle B_\bot^2\right\rangle}{6\pi}\left[\frac{1}{4}\left(
 \frac{\Gamma_2+\Gamma_{\mathrm{P1}}}{\left(\Gamma_2+\Gamma_{\mathrm{P1}}\right)^2+4(\omega_0\pm\frac{D}{2}+\frac{A_z}{2})^2} +\frac{\Gamma_2+\Gamma_{\mathrm{P1}}}{\left(\Gamma_2+\Gamma_{\mathrm{P1}}\right)^2+4(\omega_0\pm\frac{D}{2}-\frac{A_z}{2})^2}\right)\right.\nonumber\\
 &&\,\,\,\,\,\,\,\,\,\,\,\,\,\,\,\,\,\,\,\,+  \frac{3}{4}\left(\frac{\Gamma_2+\Gamma_{\mathrm{P1}}}{\left(\Gamma_2+\Gamma_{\mathrm{P1}}\right)^2+4(\omega_0\pm\frac{D}{2}+\frac{a_z}{2})^2}
 +\frac{\Gamma_2+\Gamma_{\mathrm{P1}}}{\left(\Gamma_2+\Gamma_{\mathrm{P1}}\right)^2+4(\omega_0\pm\frac{D}{2}-\frac{a_z}{2})^2}\right)\nonumber\\
&&\left. \,\,\,\,\,\,\,\,\,\,\,\,\,\,\,\,\,\,\,\,+\frac{\Gamma_2+\Gamma_{\mathrm{P1}}}{\left(\Gamma_2+\Gamma_{\mathrm{P1}}\right)^2+4(\omega_0\pm\frac{D}{2})^2}\right]\nonumber\\
&&+\frac{\left\langle B_\|^2\right\rangle}{4\pi}\left[\frac{1}{4}\left(
 \frac{\Gamma_2+\Gamma_{\mathrm{P1}}}{\left(\Gamma_2+\Gamma_{\mathrm{P1}}\right)^2+4(\omega_0\pm\frac{D}{2}+\frac{A_z}{4})^2} +\frac{\Gamma_2+\Gamma_{\mathrm{P1}}}{\left(\Gamma_2+\Gamma_{\mathrm{P1}}\right)^2+4(\omega_0\pm\frac{D}{2}-\frac{A_z}{4})^2}\right)\right.\nonumber\\
 &&\left.\,\,\,\,\,\,\,\,\,\,\,\,\,\,\,\,\,\,\,\,+  \frac{3}{4}\left(\frac{\Gamma_2+\Gamma_{\mathrm{P1}}}{\left(\Gamma_2+\Gamma_{\mathrm{P1}}\right)^2+4(\omega_0\pm\frac{D}{2}+\frac{a_z}{4})^2}
 +\frac{\Gamma_2+\Gamma_{\mathrm{P1}}}{\left(\Gamma_2+\Gamma_{\mathrm{P1}}\right)^2+4(\omega_0\pm\frac{D}{2}-\frac{a_z}{4})^2}\right)\right],
\end{eqnarray}
where the effective couplings, $\left\langle B_\bot^2\right\rangle$ and $\left\langle B_\|^2\right\rangle$, are due to integration over all possible NV-P1 separations. By taking the average P1 density to be 50\,ppm, and the FID rate to be $\Gamma_2=5.0\,$MHz, we may plot the resulting field-dependent relaxation rate of the NV spin (Fig.\,\ref{TheoreticalSpectrum}).

\begin{figure}
  \centering
  \includegraphics[width=8cm]{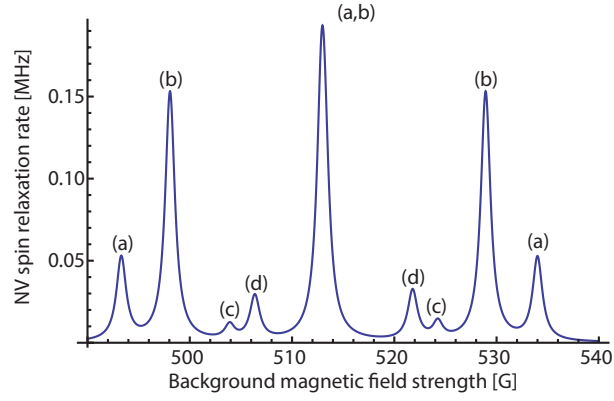}\\
  \caption{Analytic calculation of the relaxation rate of an NV centre spin placed in a 50ppm environment of P1 donor electron spins vs the strength of an external magnetic field aligned along the NV axis. (a) Peaks associated with allowed transitions of the NV centre spin due to its direct interaction with aligned P1 centre electron spins. (b) As in (a), but for the case of P1 centres not aligned along the NV axis. (c) Peaks associated with disallowed transitions of the NV spin as mediated by flip-flop dynamics between the on-axis P1 electron and nuclear spins. (d) As in (c), but for the case of off-axis P1 centres.}\label{TheoreticalSpectrum}
\end{figure}

\end{widetext}


\end{document}